\documentclass[12pt,english]{article}
\usepackage[T1]{fontenc}
\usepackage[latin1]{inputenc}
\usepackage{color}
\usepackage{graphicx}

\makeatletter

\providecommand{\tabularnewline}{\\}

\usepackage[T1]{fontenc}
\usepackage{geometry}
\usepackage{graphics}
\makeatletter

\usepackage{babel}
\makeatother
\begin{document}

\title{\textcolor{black}{$B_{c}$} photoproduction at HERA collider%
\footnote{The short talk given at Weizmann Institute Workshop on HEAVY QUARK
PHYSICS AT HERA II (Rehovot, Israel, October 19-22, 2003). %
}}

\author{A.V.Berezhnoy %
\footnote{Scobeltsyn Institute for Nuclear Physics of Moscow State University,
Moscow, Russia.%
}, A.K.Likhoded %
\footnote{Institute for High Energy Physics, Protvino, Russia.%
}}

\date{}

\maketitle
\begin{abstract}
The $B_{c}$ photoproduction at HERA has been considered in the frame
work of the pQCD. The estimated value of the total production cross
section of $B_{c}^{+}$, $B_{c}^{*+}$, $B_{c}^{-}$ and $B_{c}^{*-}$
mesons is about 5 pb. Higher excitations yield about 30\% to the total
cross section. Thus for integrated HERA luminosity 1 $\textrm{fb}^{-1}$
about $7\cdot10^{3}$ $B_{c}$ mesons will be produced. 
\end{abstract}

\section{Calculation technique}

The main features of the $ep$-production of $B_{c}$in the frame
work of pQCD had been discussed in our previous work\cite{previous}.
The dominant subprocess $\gamma g\rightarrow B_{c}+X$ is described
by 24 Feynman diagrams. We suppose that the amplitude of the production
process can be expressed through the amplitude $T$ of hard production
of four heavy quarks $b\bar{b}c\bar{c}$ followed by the soft fusion
of $\bar{b}$ and $c$ quarks into $B_{c}$ meson due to the meson
wave function $\Psi$: \[
A_{\gamma g\rightarrow B_{c}+b+\bar{c}}\sim\int T_{\gamma g\rightarrow b\bar{b}c\bar{c}}^{\textrm{hard}}(p_{i},\vec{q})\cdot\tilde{\Psi}_{q}^{*}(\vec{q})d^{3}q,\]
 where $p_{i}$ are four momenta of $B_{c}$, $b$ and $\bar{c}$;
$\vec{q}$ is the three momentum of $\bar{b}$ quark in the $B_{c}$
meson rest frame.

One can neglect the $\vec{q}$ dependence in $T_{\gamma g\rightarrow b\bar{b}c\bar{c}}^{\textrm{hard}}$
and get for $S$ wave states production the following equation: 

\[
A_{\gamma g\rightarrow B_{c}+b+\bar{c}}\sim\Psi(0)\cdot T_{\gamma g\rightarrow b\bar{b}c\bar{c}}^{\textrm{hard}}(p_{i},0),\]
 where $\Psi(0)$ is the value of $B_{c}$ wave function value at
origin (coordinate space).

In such approach $\bar{b}$ and $c$ quarks in $B_{c}$ have the same
velocities. Thus: \[
p_{\bar{b}}=\frac{m_{b}}{M_{B_{c}}}P_{B_{c}},\]
 \[
p_{c}=\frac{m_{c}}{M_{B_{c}}}P_{B_{c}}.\]

If one need to get the yields of $B_{c}$ and $B_{c}^{*}$ separately,
the product of spinors $v_{\bar{b}}\bar{u}_{c}$, corresponding to
the $\bar{b}$ and $c$ quarks in the $T_{\gamma g\rightarrow b\bar{b}c\bar{c}}^{\textrm{hard}}$
amplitude, should be substituted by the projection operator \[
{\mathcal{P}}(\Gamma)=\sqrt{M}\left(\frac{\frac{m_{b}}{M}\hat{P}_{B_{c}}-m_{b}}{2m_{b}}\right)\Gamma\left(\frac{\frac{m_{c}}{M}\hat{P}_{B_{c}}+m_{c}}{2m_{c}}\right),\]
 where $\Gamma=\gamma^{5}$ for $s=0$, or $\Gamma=\hat{\varepsilon}^{*}(P_{B_{c}},s_{z})$
for $s=1$, where $\varepsilon(P_{B_{c}},s_{z})$ is the polarization
vector for the spin-triplet state.

Nevertheless, one can easily show that $\Gamma$ can be expressed
through the $v_{\bar{b}}$ and $\bar{u}_{c}$:

\[
{\mathcal{P}}(\gamma^{5})=\sqrt{\frac{2M}{2m_{b}2m_{c}}}\frac{1}{\sqrt{2}}\{ v_{b}(p_{b},+)\bar{u}_{c}(p_{c},+)-v_{b}(p_{b},-)\bar{u}_{c}(p_{c},-)\},\]

\[
{\mathcal{P}}(\hat{\varepsilon}^{*}(P_{B_{c}},-1))=\sqrt{\frac{2M_{B_{c}}}{2m_{b}2m_{c}}}v_{b}(p_{b},+)\bar{u}_{c}(p_{c},-),\]

\[
{\mathcal{P}}(\hat{\varepsilon}^{*}(P_{B_{c}},0))=\sqrt{\frac{2M_{B_{c}}}{2m_{b}2m_{c}}}\frac{1}{\sqrt{2}}\{ v_{b}(p_{b},+)\bar{u}_{c}'(p_{c},+)+v_{b}(p_{b},-)\bar{u}_{c}(p_{c},-)\},\]

\[
{\mathcal{P}}(\hat{\varepsilon}^{*}(P_{B_{c}},+1))=\sqrt{\frac{2M_{B_{c}}}{2m_{b}2m_{c}}}v_{b}(p_{b},-)\bar{u}_{c}(p_{c},+).\]

The more detailed description of the calculation technique one can
find in the papers \cite{technique}, where the gluonic and photonic
$B_{c}$ production cross sections had been calculated (see also \cite{others}).

To get the cross section of $B_{c}$ production in $ep$ collision
one should covert the cross section of the subprocess $\gamma g\rightarrow B_{c}+X$
with the distribution function of gluon in the initial proton (we
use CTEQ6 parametrization of PDF \cite{PDF}) and Weizs\"{a}cker-Williams
photon function.

\section{Results}

The parameter values have been chosen as follows:

\[
\alpha_{s}=0.2,\]
 \[
m_{c}=1.5\,\,\textrm{GeV},\]
 \[
m_{b}=4.8\,\,\textrm{GeV}\]
 \[
|\Psi(0)|^{2}=0.116\,\,\textrm{GeV}^{3}.\]

The total production cross section of $B_{c}^{+}$, $B_{c}^{*+}$,
$B_{c}^{-}$ and $B_{c}^{*-}$ mesons is about 5 pb. Higher excitations
yield about 30\% to the total cross section. Thus for integrated HERA
luminosity 1 $\textrm{fb}^{-1}$ about $7\cdot10^{3}$ $B_{c}$ mesons
will be produced. It is rather small amount of $B_{c}$ to find these
particles at HERA, because of small branching ratios of $B_{c}$ decay
mode, which are interesting for the extraction of $B_{c}$ signal.
Nevertheless, one can hope to find these particles at HERA collider. 

\begin{table}

\caption{The branching ratios of exclusive $B_{c}$ decay modes \cite{table}.}

\begin{center}\begin{tabular}{|c|c|c|c|}
\hline 
$B_{c}$ decay mode &
 BR, \%&
 $B_{c}$ decay mode &
 BR, \%\tabularnewline
\hline
$\psi l^{+}\nu_{l}$&
 2.5 &
 $\eta_{c}l^{+}\nu_{l}$&
 1.2 \tabularnewline
 $B_{s}^{*}l^{+}\nu_{l}$&
 6.2 &
 $B_{s}l^{+}\nu_{l}$&
 3.9 \tabularnewline
 $\psi\pi^{+}$&
 0.2 &
 $\eta_{c}\pi^{+}$&
 0.2 \tabularnewline
 $B_{s}^{*}\pi^{+}$&
 5.2 &
 $B_{s}\pi^{+}$&
 5.5 \tabularnewline
 $\psi\rho^{+}$&
 0.6 &
 $\eta_{c}\rho^{+}$&
 0.5 \tabularnewline
 $B_{s}^{*}\rho^{+}$&
 22.9 &
 $B_{s}\rho^{+}$&
 11.8  \tabularnewline
\hline
\end{tabular}\end{center}
\end{table}

\section{Remarks}

The most popular model of $B$ meson production is the fragmentation
one. In accordance to this mechanism at large $p_{T}>p_{T}^{\textrm{f}}$
quark $b$ quark is produced in the hard process followed by the soft
process of heavy quark fragmentation. For example: $\gamma g\rightarrow b\bar{b}$
followed by $b\rightarrow B$. The meson production cross section
in the frame work of fragmentational model have a very simple form:
\[
\frac{d\sigma(\gamma g\rightarrow B+X)}{dz}=\sigma(\bar{b}b)\cdot D_{b\rightarrow B}(z),\]
 where $z=2E_{B_{c}}/\sqrt{s_{\gamma g}}$.

But can we estimate the numerical value \char`\"{}large\char`\"{}
$p_{T}^{\textrm{f}}$?

For $B_{c}$ production we can do it in the frame of model discussed
in Chapter~1 and 2 of this paper: \[
p_{T}^{\textrm{f}}\sim30\,\,{\textrm{GeV}}\sim5M_{B_{c}}\]

At $p_{T}<5M_{B_{c}}$ nonframentational terms (recombination terms)
give the dominant contribution into $B_{c}$ production cross section.

\section*{Acknowledgements}

We would like to thank L. Gladilin for the fruitful discussion. The
work was supported in part by grants of RF Education Ministry E00-3.3-62,
CRDF M0-011-0, Russian Scientific School 1303.2003.

\begin{figure}
\centering

\includegraphics{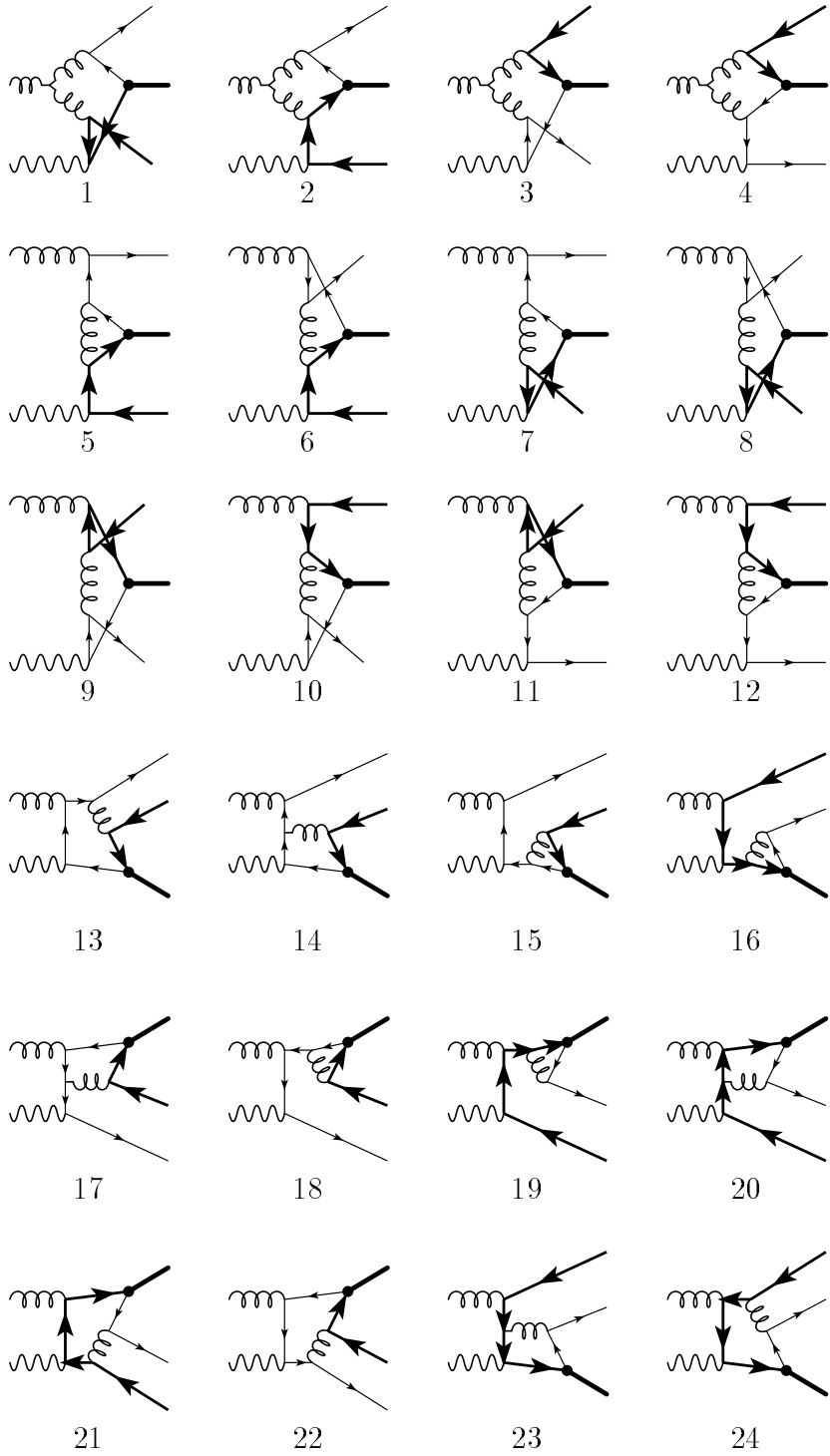}

\caption{Feynman diagrams for the process $\gamma g\rightarrow B_{c}+X$}
\end{figure}

\begin{figure}
$\frac{d\sigma_{B_{c}}}{p_{T}}$, pb/GeV

\centering

\includegraphics[%
  width=0.85\textwidth]{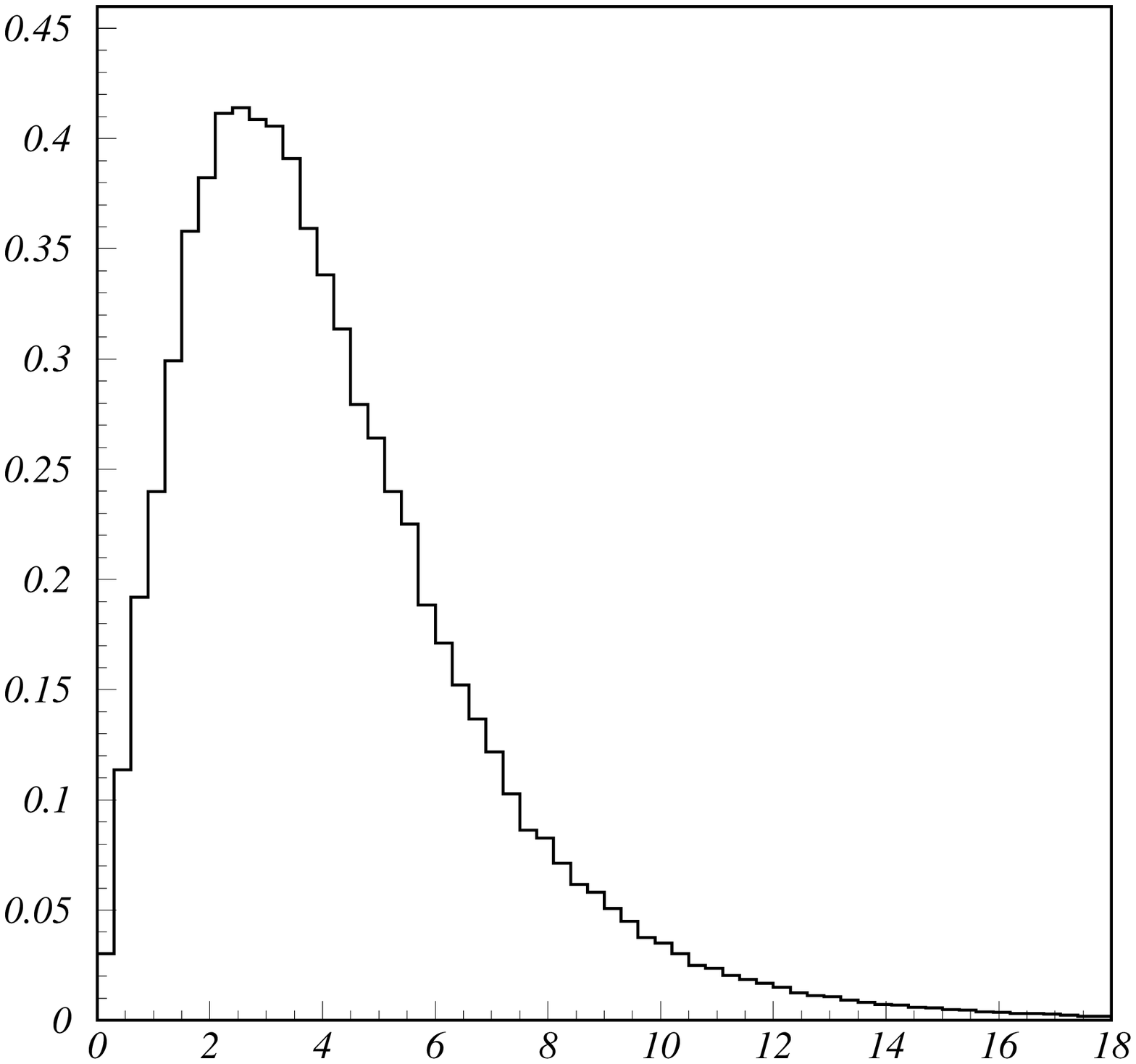}

\hfill $p_{T}$, GeV

\caption{The cross section distributions on $p_{T}$ for $B_{c}$ photoproduction
at HERA}
\end{figure}

\begin{figure}
${\displaystyle \frac{d\sigma_{B_{c}}}{d\eta}}$, pb

\centering

\includegraphics[%
  width=0.85\textwidth]{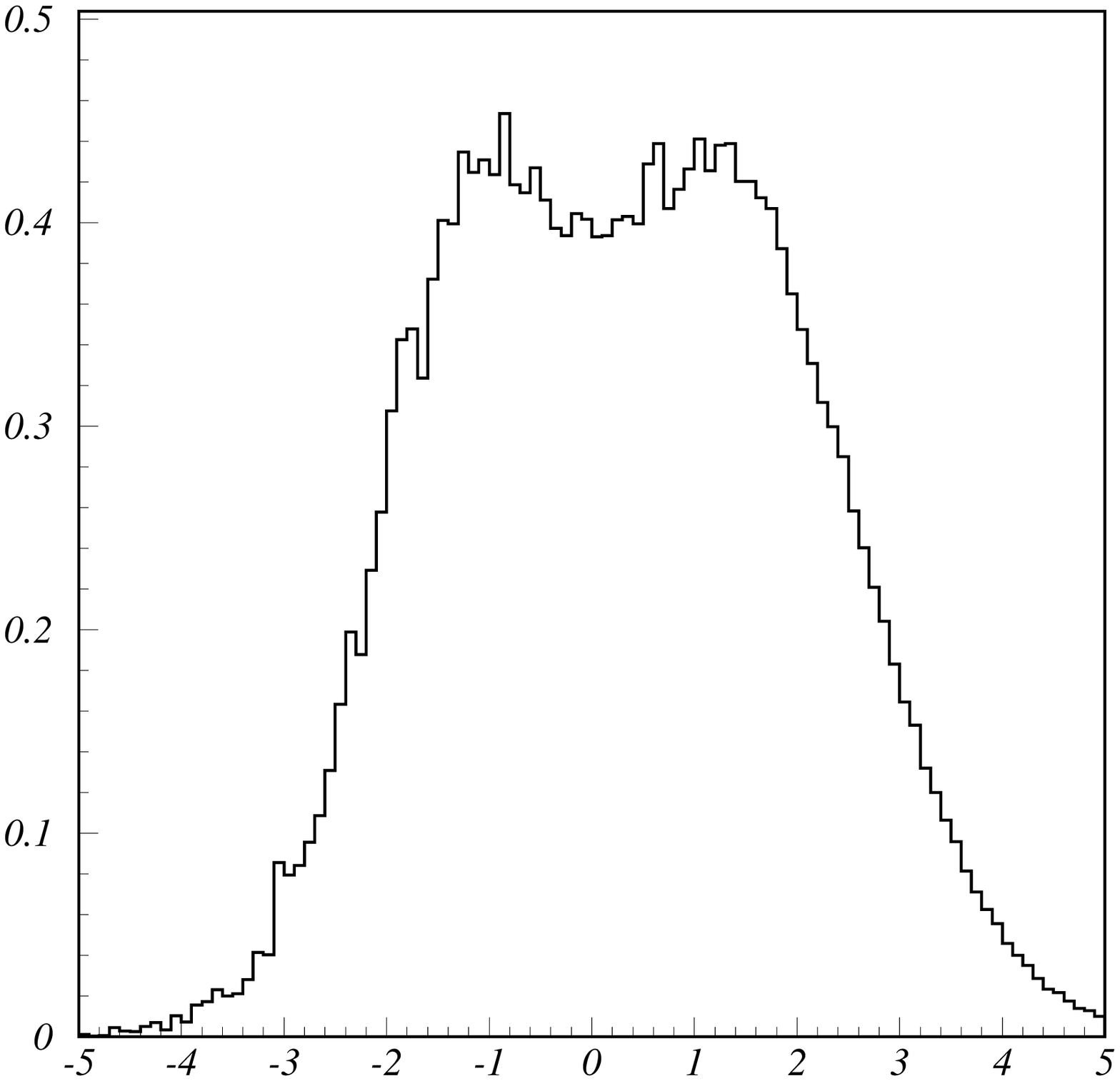}

\hfill $\eta$

\caption{The cross section distributions on $\eta$ for $B_{c}$ photoproduction
at HERA}
\end{figure}

\begin{figure}
${\displaystyle \frac{d\sigma_{B_{c}}}{d\hat{s}_{\gamma g}}}$, pb

\centering

\includegraphics[%
  width=0.85\textwidth]{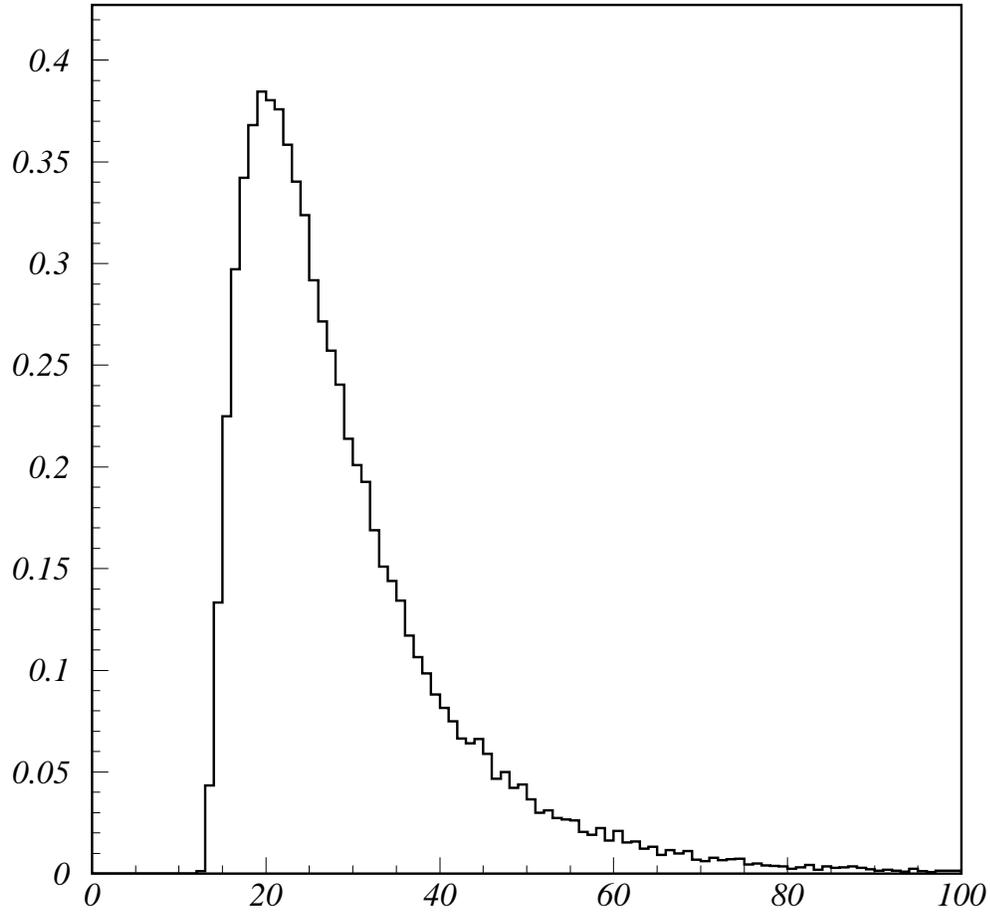} \hfill $\hat{s}_{\gamma g}$

\caption{The cross section distribution on $\hat{s}_{\gamma g}$ for $B_{c}$
photoproduction at HERA}
\end{figure}

\end{document}